\documentclass[prl,
showpacs,
twocolumn,
floats,
10pt,
aps,
citeautoscript,
longbibliography,
superscriptaddress]{revtex4-2}

\usepackage{natbib}
\usepackage{placeins}
\usepackage{lipsum}
\usepackage{xcolor}
\usepackage[normalem]{ulem}
\usepackage{comment}
\usepackage{tabularx}
\usepackage{graphicx}
\usepackage{dcolumn}
\usepackage{bm}
\usepackage{ulem}

\usepackage{bbm}
\usepackage{blindtext}
\usepackage{graphics}
\usepackage{verbatim}   
\usepackage{amsfonts}
\usepackage{amsmath}
\usepackage{amssymb}
\usepackage{adjustbox}
\usepackage{microtype} 
\allowdisplaybreaks 
\usepackage{xspace} 
\usepackage{multirow} 
\usepackage{tabstackengine}
\setstackEOL{\cr}

\newcommand{\Fig}[1]{Fig.~\ref{#1}}

\def\mi{\mathrm{i}}   
\def\me{\mathrm{e}}   
\newcommand{\mr}[1]{\mathrm{#1}}

\newcommand{\bra}[1]{\ensuremath{\langle #1 |}}
\newcommand{\ket}[1]{\ensuremath{| #1 \rangle}}

\newcommand{\expvalue}[2]{\ensuremath{\langle #1 | #2 | #1 \rangle}}

\newcommand*{\ndots}{\kern-0.075em.\kern-0.05em.\kern-0.05em.}  
\newcommand*{\nidots}{.\kern-0.05em.\kern-0.05em.} 
\newcommand*{\ncdots}{\kern-0.15em\cdot\kern-0.2em\cdot\kern-0.2em\cdot\kern-0.15em}   

\NewDocumentCommand{\doubleI}{O{}}{\mathbbm{1}_{#1}}
\NewDocumentCommand{\doubleIb}{O{}}{{\overline{\mathbbm{1}}_{#1}}}
\NewDocumentCommand{\doubleIk}{O{}}{\mathbbm{1}^\ks_{\! #1}}
\NewDocumentCommand{\doubleId}{O{}}{\mathbbm{1}^\ds_{\! #1}}
\NewDocumentCommand{\doubleIp}{O{}}{\mathbbm{1}^\ps_{\! #1}}
\NewDocumentCommand{\doubleV}{O{}}{\mathbbm{V}_{\! #1}}
\NewDocumentCommand{\doubleVk}{O{}}{\mathbbm{V}^\ks_{\! #1}}
\NewDocumentCommand{\doubleVd}{O{}}{\mathbbm{V}^\ds_{\! #1}}
\NewDocumentCommand{\doubleVp}{O{}}{\mathbbm{V}^\ps_{\! #1}}
\NewDocumentCommand{\doublev}{o}{{\mathbbm{v}_{#1}}}
\NewDocumentCommand{\doubleVb}{o}{{\overline{\mathbbm{V}}_{\! #1}}}
\NewDocumentCommand{\doubleVt}{o}{{\widetilde{\mathbbm{V}}_{\! #1}}}
\NewDocumentCommand{\doubleVh}{o}{\widehat{{\mathbbm{V}}_{\! #1}}}
\NewDocumentCommand{\doubleW}{o}{\mathbbm{W}_{\! #1}}
\NewDocumentCommand{\doubleWk}{o}{\mathbbm{W}^\ks_{\! #1}}
\NewDocumentCommand{\doubleWd}{o}{\mathbbm{W}^\ds_{\! #1}}
\NewDocumentCommand{\doubleWb}{o}{{\overline{\mathbbm{W}}_{\! #1}}}
\NewDocumentCommand{\doubleWt}{o}{{\widetilde{\mathbbm{V}}_{\! #1}}}
\NewDocumentCommand{\doubleWh}{o}{{\widehat{\mathbbm{V}}_{\! #1}}}

\newcommand{\LMUMunich}{Department of Physics and Arnold Sommerfeld Center for Theoretical Physics (ASC), Ludwig-Maximilians-University Munich,
Theresienstr. 37, D-80333 Munich, Germany}

\newcommand{\MCQST}{Munich Center for Quantum Science and Technology, Schellingstr. 4, D-80799 Munich, Germany}

\newcommand{\Regensburg}{Institut für Theoretische Physik, Universität Regensburg, D-93035 Regensburg, Germany}

\definecolor{darkgreen}{rgb}{0,0.5,0}
\definecolor{purple}{rgb}{0.6,0,0.5}
\definecolor{orange}{rgb}{1,0.5,0}
\definecolor{darkred}{rgb}{.7,0,0}
\definecolor{darkblue}{rgb}{0,0,.6}
\definecolor{grey}{rgb}{.6,.6,.6}
\definecolor{dimgreen}{rgb}{0.2,0.7,0.2}
\definecolor{brightgreen}{rgb}{0.5020, 1, 0}

\newcommand{\jvdomit}[1]{}

\usepackage{multirow}
\usepackage{physics}
\usepackage{hyperref}
\hypersetup{colorlinks=true,breaklinks,linkcolor=blue,urlcolor=blue,citecolor=blue}

\usepackage{graphicx}
\usepackage{dcolumn}
\usepackage{bm}

\usepackage[caption=false]{subfig}
\usepackage{braket}

\begin{document}

\preprint{}

\title{Magnetic polarons at finite temperature: One-hole spectroscopy study}


\author{Toni Guthardt}
\affiliation{\LMUMunich}

\author{Markus Scheb}%
\affiliation{\LMUMunich}

\author{Jan von Delft}
\affiliation{\LMUMunich}
\affiliation{\MCQST}

\author{Annabelle Bohrdt}
\affiliation{\MCQST}
\affiliation{\Regensburg}

\author{Fabian Grusdt}
\affiliation{\LMUMunich}
\affiliation{\MCQST}



\date{\today}

\begin{abstract}
The physics of strongly correlated fermions described by Hubbard or $t$-$J$ models in the underdoped regime -- relevant for high-temperature superconductivity in cuprate compounds -- remains a subject of ongoing debate. In particular, the nature of charge carriers in this regime is poorly understood, in part due to the unusual properties of their spectral function. In this Letter, we present unbiased numerical results for the one-hole spectral function in a $t$-$J$ model at finite temperatures. Our study provides valuable insights into the underlying physics of magnetic (or spin-) polaron formation in a doped antiferromagnet (AFM). For example, we find how the suppression of spectral weight outside the magnetic Brillouin zone -- a precursor of Fermi arc formation -- disappears with increasing temperature, revealing nearly-deconfined spinon excitations of the undoped AFM. The pristine setting we consider can be directly explored using quantum simulators. Our calculations demonstrate that coherent quasiparticle peaks associated with magnetic polarons can be observed up to temperatures $T>J$ above the spin-exchange $J$, routinely obtained in such experiments. This paves the way for future studies of the fate of magnetic polarons in the pseudogap phase. 
\end{abstract}

\maketitle

\emph{Introduction.---}
The disappearance of antiferromagnetism underlies a variety of exotic phenomena of strongly correlated electrons, including heavy-fermion superconductivity related to Kondo couplings~\cite{Stewart1984}, or pseudogap formation in hole-doped cuprate compounds~\cite{Lee2006highTc,Keimer2015}. A natural starting point for understanding the underlying physics is to work in the strong-coupling limit, where Hubbard interactions $U \gg t$ dominate over tunneling $t$, and consider the elementary charge carriers of the doped AFM Mott insulator. These magnetic, or spin-, polarons~\cite{Bulaevskii1968,Trugman1988,kane1989motion,sachdev1989hole} correspond to the quasiparticles formed upon removing a single fermion from the Mott insulator and creating a mobile vacancy, the doped hole. They have been observed experimentally in solids~\cite{Ronning2005,Graf2007} and in neutral atom quantum simulators~\cite{Koepsell2019,Ji2021,qiao2025realization}.

Angle-resolved photoemission spectroscopy~\cite{Damascelli2003} (ARPES) studies on hole-doped cuprates provide evidence that the physics of magnetic polarons, whose ground states are well understood deep in the AFM phase~\cite{kane1989motion,sachdev1989hole,Trugman1990,Elser1990,Martinez1991,Boninsegni1991,Liu1991,Beran1996,Grusdt2019PRB,Nielsen2021,Bermes2024}, is related to the physics of Fermi arcs in the pseudogap regime. Specifically, laser-ARPES studies on few-layer copper-oxide compounds suggest that the small Fermi pockets around the nodal point $(\pi/2,\pi/2)$ associated with magnetic polarons continuously evolves into Fermi arcs with decreasing spectral weight outside the magnetic Brillouin zone (BZ) as doping is increased from $1\%$ to around $5\%$~\cite{Kurokawa2023}. This is consistent with theoretical calculations based on linear spin-wave theory~\cite{Chen2011d} and cellular dynamical mean-field theory~\cite{BacqLabreuil2025}, and motivates further studies of the distribution of spectral weight of magnetic polarons.

\begin{figure}[b]
\centering
\includegraphics[width = 0.47\textwidth]{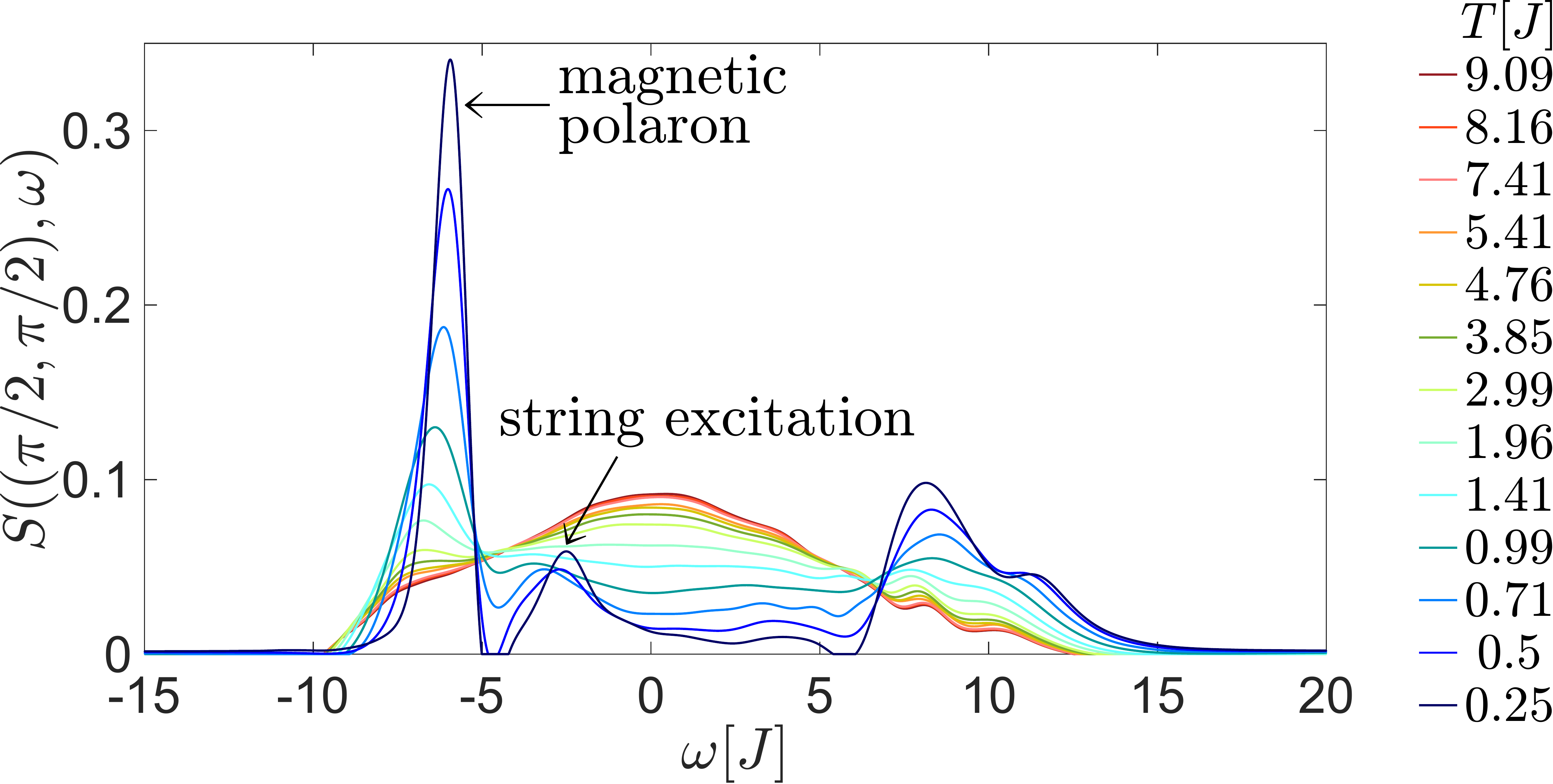}
\caption{Single-hole ARPES spectrum at the nodal point $\mathbf{k}=(\pi/2,\pi/2)$, computed for various temperatures $T$ indicated on the right. We consider a $t$-$J$ model at $t/J = 3$, on a four-leg cylinder accessible to our time-dependent MPS simulations that we combine with purification schemes to obtain finite-$T$ spectra. At low energies, around $\omega \lesssim - 6 J$, a well-defined quasiparticle peak associated with magnetic polaron formation is visible up to fairly high temperatures ($T\simeq 1.4 J$). }
\label{FigSpectralCut}
\end{figure}

In this Letter, we present results from unbiased numerical simulations of the one-hole spectral function in the $t$-$J$ model at variable temperatures, based on matrix product states (MPS)~\cite{Hauschild2019disent,Paeckel2019a,Li2024TDVP}. First we reveal stable magnetic polaron features in the spectrum up to temperatures $T >J$ well above the spin-exchange $J$, see Fig.~\ref{FigSpectralCut}. This is remarkable, given the quick demise of AFM correlations beyond nearest neighbor sites at these high temperatures. Second, when temperature is increased we report on the appearance of low-energy spectral weight outside the magnetic BZ, where the signal is strongly suppressed at low temperatures in a precursor of Fermi-arc formation. We explain these features in terms of thermally excited, nearly-deconfined spinons~\cite{Piazza2015} in the parent AFM.

The spectral properties of magnetic polarons in the $t$-$J$ model have been investigated in great detail at $T=0$. Initial work was based on the linear-spin wave approximation~\cite{kane1989motion,sachdev1989hole}, and combined with the self-consistent Born approximation (SCBA) to predict the shape of the one-hole spectrum~\cite{SchmittRink1988,Liu1991,Martinez1991,Liu1992,Nielsen2022}. While the SCBA has been validated on the level of the linear spin-wave Hamiltonian~\cite{Diamantis2021}, unbiased numerical studies of the one-hole spectrum in the more fundamental $t$-$J$ model, including exact diagonalization~\cite{Leung1995,Leung1996}, diagrammatic Monte Carlo~\cite{Brunner2000,Mishchenko2001} and MPS~\cite{Bohrdt2020}, have revealed discrepancies at higher energies. These have been traced back to interactions among spin-waves~\cite{Wrzosek2021,Bermes2024}, which lead to the disappearance of string-excitations of magnetic polarons~\cite{Bulaevskii1968,Shraiman1988,Dagotto1990,Liu1992,Manousakis2007,Grusdt2018PRX} beyond the first vibrational peak~\cite{Mishchenko2001,Bohrdt2020}. The finite-temperature one-hole spectrum remains much less studied, with the notable exception of a linear-spin wave analysis~\cite{Kar2008}.

One of the most intriguing features of the $T=0$ one-hole spectrum, not captured by SCBA, is a strong suppression of spectral weight outside the magnetic BZ up to energies on the order of $t$ above the ground state when $t > J$~\cite{Brunner2000,Bohrdt2020}. This phenomenon has been argued~\cite{Bohrdt2020} to be related to the formation of Fermi-arcs, through features in the spectrum associated with nearly-deconfined spinons. Our analysis of finite temperature spectra in this Letter provides direct evidence for this picture. 

\emph{Model and method.---}
Throughout this work we consider zero or one hole in the $t$-$J$ Hamiltonian,
\begin{equation}
\hat{H} = -t\sum_{\braket{\mathbf{i},\mathbf{j}},\sigma}\mathcal{\hat{P}}(\hat{c}^{\dagger}_{\mathbf{i},\sigma}\hat{c}_{\mathbf{j},\sigma} + {\rm h.c.})\mathcal{\hat{P}} + J \sum_{\braket{\mathbf{i},\mathbf{j}}} \left( \mathbf{\hat{S}_i} \cdot \mathbf{\hat{S}_j}-\frac{\hat{n}_\mathbf{i} \hat{n}_\mathbf{j}}{4} \right),
\label{eq:tjmodel}
\end{equation}
where $\mathcal{\hat{P}}$ projects onto states with at most one fermion $\hat{c}_{\mathbf{j},\sigma}$ per site, i.e. $\hat{n}_{\mathbf{j}}=\sum_\sigma \hat{c}^\dagger_{\mathbf{j},\sigma} \hat{c}_{\mathbf{j},\sigma} \leq 1$, and $\hat{\mathbf{S}}_{\mathbf{j}}$ is the spin operator at site $\mathbf{j}$. This model has been proposed to describe high-$T_c$ cuprate superconductors~\cite{Zhang1988}, captures key properties of their phase diagrams~\cite{Jiang2021} and can be obtained, up to a correlated hole-hopping term $\propto J=4 t^2/U$, as the low-energy limit of the Hubbard model.

In addition to their (approximate) realizations in solids, pristine implementations of $t$-$J$~\cite{Gorshkov2011tJ,Homeier2024,Carroll2024,Qiao2025} and Fermi-Hubbard models~\cite{Esslinger2010,Tarruell2018,Bohrdt2021PWA} with tunable model parameters have been achieved in neutral atom quantum simulators, as well as in digital platforms~\cite{Barends2015,Evered2025}. In all of these experimental settings, the momentum-resolved spectral function that we study in this Letter can be measured, using ARPES in solids or neutral-atom incarnations of the latter in quantum simulators~\cite{Kollath2007,Dao2007,Dao2009,Torma2016,Bohrdt2018,Brown2019,Homeier2024}.

We calculate the one-hole spectral function, $S(\mathbf{k},\omega) = - \Im{A(\mathbf{k},\omega)} / \pi$, from the Fourier transformation, 
\begin{equation}
    A(\mathbf{k},\omega) = \int_{-\infty}^{\infty} d\tau ~ \me^{- \mi \omega \tau} \sum_{\mathbf{j}} \me^{- \mi \mathbf{k} \mathbf{j}} C_{\mathbf{0},\mathbf{j}}(\tau),
\label{eqAkwFourierTrafo}
\end{equation}
of a space and time-dependent correlation function $C_{\mathbf{i},\mathbf{j}}(\tau)$. The latter can be directly computed using MPS as the time-evolution of an initial thermal state of the AFM with a single hole created at site $\mathbf{i}$ (we set $\hbar = 1$),
\begin{equation}
    C_{\mathbf{i},\mathbf{j}}(\tau)=-\mi \sum_{\sigma} \expvalue{\psi_{\mr{equil}}^{\mr{puri}}}{\me^{\mi \hat{H}\tau} \hat{c}^{\dagger}_{\mathbf{j},\sigma} \me^{-\mi \hat{H}\tau} \hat{c}_{\mathbf{i},\sigma}}.
\label{time_dependent_correl}
\end{equation}
Here $\ket{\psi_{\mr{equil}}^{\mr{puri}}}$ is the finite temperature purified state of the undoped AFM in equilibrium. 

Our simulations start by calculating $\ket{\psi_{\mr{equil}}^{\mr{puri}}}$ on a cylinder with length $L_x = 18$ and width $L_y=4$. We use the density matrix renormalization group~\cite{White1992Nov,White1993Oct} in the language of MPS~\cite{Schollwoeck2011dmrg}, adapted to finite temperatures via a purification scheme~\cite{Nocera2016puri,Feiguin2005puri,Feiguin2010puri} and enhanced by the use of disentanglers~\cite{Hauschild2019disent}. We simulate the time-evolution in Eq.~\eqref{time_dependent_correl} by combining two MPS-based algorithms~\cite{paeckel2019timeevol}: To capture the initial spreading of entanglement across the MPS we begin with a single step of the more expensive, global Krylov scheme~\cite{Garcia-Ripoll2006Dec,Dargel2012May,Wall2015}. Then we switch to the local, but less expensive time-dependent-variational-principle algorithm~\cite{Haegeman2011TDVP,Haegeman2016TDVP}. This procedure is improved by the use of a backwards-time-evolution scheme \cite{Karrasch2012anc,Kennes2016anc,Karrasch2013anc}, which allows to reach longer times without further approximations. In addition, we make use of controlled bond expansion~\cite{Gleis2023CBE,Li2024TDVP}, which effectively performs two-site optimization at one-site cost. Symmetries in tensor network computations were exploited using the QSpace tensor library~\cite{Weichselbaum2012,Weichselbaum2020,Weichselbaum2024}. Finally, in order to extend the time window that we can use for the integration in Eq.~\eqref{eqAkwFourierTrafo}, we employ linear prediction \cite{leland1989digital} and multiply the correlation function $C_{\mathbf{0},\mathbf{j}}(t)$ with a Gaussian envelope~\cite{verresen2018quantum}.

\begin{figure*}[t]
\centering
\includegraphics[width = \textwidth]{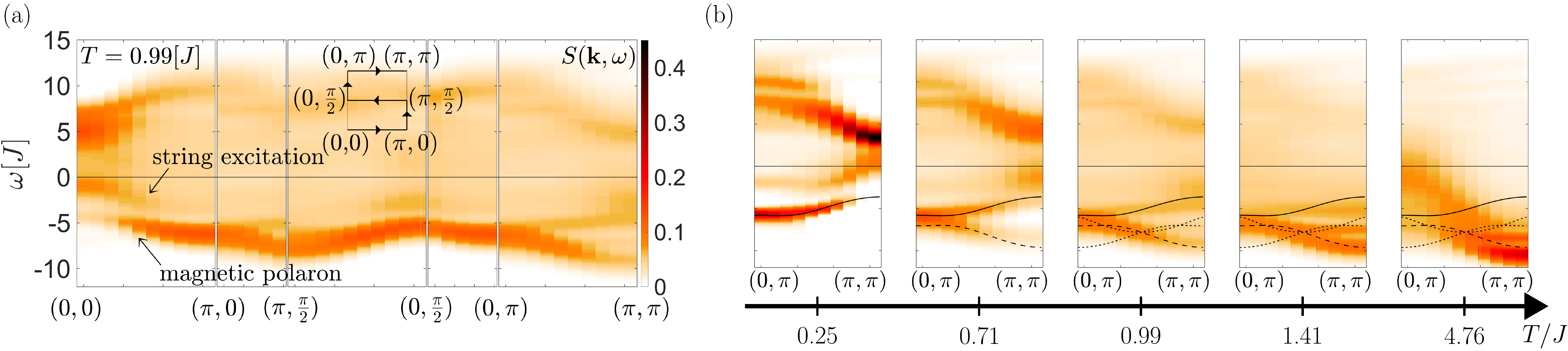}
\caption{Finite temperature ARPES spectra, computed for $t/J=3$ on a four-leg cylinder. (a) We show the single-hole spectrum at $T=0.99 J$ along the S-shaped cut through the BZ shown in the inset. At low energies the characteristic spectral signatures of magnetic polarons remain visible. (b) We show the evolution of the spectrum as temperature $T/J$ is increased, along a cut from $(0,\pi)$ to $(\pi,\pi)$. Around $(\pi,\pi)$, a pronounced spectral gap for $\omega \lesssim 0$ is visible at the lowest temperatures which gradually fills up as $T$ is increased. The predicted low-energy thermal spinon lines (dashed and dotted lines) for this four-leg cylinder are also indicated, along with the approximate magnetic polaron dispersion (solid lines), as discussed in the main text.}
\label{FigFullSpectra}
\end{figure*}

\emph{Quasiparticle properties at finite $T$.---}
We begin the discussion of our results by focusing on the quasiparticle peak associated with the formation of a magnetic polaron. Fig.~\ref{FigSpectralCut} shows spectral cuts $S(\mathbf{k},\omega)$ for different temperatures at the nodal point, $\mathbf{k}=(\pi/2,\pi/2)$, where the minimum of the magnetic polaron dispersion is located. We consider the case $t/J=3$, but obtain similar results for $t/J=1$ and $5$, see supplements. For the lowest temperatures, $T=0.25 J$, we observe a sharp quasiparticle peak around $\omega\approx -6 J$ whose width is Fourier-limited by the finite time of our accessible time-dependent MPS calculations of Eq.~\eqref{time_dependent_correl}.

With increasing temperature, up to $T \lesssim 2 J$, the magnetic polaron peak remains clearly visible. In this regime its energy shifts to lower frequency, reaching $\omega \approx -7 J$ at $T \approx 2 J$. Moreover the peak broadens and its width is temperature- instead of Fourier-limited. This behavior is typical for polaronic models, independent of the details of the model~\cite{Guenther2018}. At temperatures above $T \gtrsim 2 J$, the coherent quasiparticle feature disappears, making way for a low-energy shoulder which finally also dissolves above $T \gtrsim 5 J$. At that point, the entire spectrum becomes featureless and broad. The associated energy scale is consistent with the magnetic polaron bandwidth of $W \approx 2 J$~\cite{Martinez1991,Beran1996,Grusdt2019PRB}. Similar behavior is found at other momenta within the magnetic BZ, see Fig.~\ref{FigFullSpectra}(a). 

For the lowest temperatures the first string excitation peak of the polaron~\cite{Brunner2000,Mishchenko2001,Manousakis2007,Bohrdt2020}, at $\Delta \omega \approx 3.5 J$ above the ground state, is also clearly visible in Fig.~\ref{FigSpectralCut}. Like the main quasiparticle peak, this feature broadens and shifts to lower frequencies before it vanishes at temperatures $T\approx J$, somewhat before the magnetic polaron ground state disappears. Closer to the center of the BZ around $\mathbf{k}=(0,0)$, this first string excitation remains visible up to slightly higher temperatures, see Fig.~\ref{FigFullSpectra}(a).

The robustness of the magnetic polaron features that we find upon increasing temperature to between $J$ and $2 J$ is remarkable, given that the undoped parent AFM in two dimensions has no long-range order at any $T>0$. The AFM correlation length diverges exponentially with $1/T$ and reaches a few lattice sites only below $T \lesssim 0.6 J$~\cite{Mazurenko2017}. On the other hand, the nearest-neighbor spin-spin correlations of the parent AFM become sizable already around $T \lesssim J$. This suggests that the presence of local spin correlations is essential for seeing spectral signatures of magnetic polarons, rather than long-range AFM order.

\emph{Low-energy spectral weight.---}
Next we discuss the evolution of low-energy spectral weight around $\mathbf{k}=(\pi,\pi)$ outside the magnetic BZ as temperature is increased. Our numerical results are shown in Fig.~\ref{FigFullSpectra}(b), focusing on a cut from $(0,\pi)$ to $(\pi,\pi)$. In agreement with calculations at $T=0$~\cite{Bohrdt2020}, at low temperatures $T \ll J$ we observe a nearly complete suppression of spectral weight below $\omega \lesssim 0$ at $\mathbf{k}=(\pi,\pi)$. Since the undoped parent AFM ground state breaks the translational symmetry, single-hole eigenstates at $\mathbf{k}=(\pi,\pi)$ exist at the same eigenenergies as at $\mathbf{k}=(0,0)$, where spectral weight is observed down to $\omega_{\rm min} \approx - t$. I.e., all magnetic polaron states at $\mathbf{k}=(\pi,\pi)$ with energies below $\mathcal{O}(t)$ above the ground state have negligible spectral weight. We confirm this picture by our calculations of the one-hole spectrum at $t/J=5$, see supplements, where an even larger spectral gap $\approx 10 J$ is found at $\mathbf{k}=(\pi,\pi)$ at low temperature. 

This behavior can be interpreted~\cite{Bohrdt2020} as a precursor of Fermi-arc formation observed in hole-doped cuprates in the pseudogap regime, where a sharp drop of spectral weight across the edge of the magnetic BZ is found~\cite{Shen2005,Kurokawa2023}. A similar feature is well-known to arise in the one-hole spectrum of a one-dimensional (1D) spin-chain~\cite{Kim2006}, where it is explained by a combination of spin-charge separation and the description of the one-dimensional Heisenberg AFM as a spinon Fermi sea~\cite{Szczepanski1990,Eder1997,Bannister2000}. However, exact numerical studies in 1D systems found that this feature disappears upon increasing temperature~\cite{Bohrdt2018}, which can be traced back to thermal excitations of spinon states above the spinon Fermi surface~\cite{Bohrdt2018,Schuckert2021a}. Now we report on a similar effect in two dimensional systems.

Upon increasing temperature beyond $T\gtrsim 0.7 J$, in Fig.~\ref{FigFullSpectra}(b) we indeed observe the gradual re-emergence of low-energy spectral weight at $\mathbf{k}=(\pi,\pi)$. On the one hand, the magnetic polaron branch  featuring a dispersion maximum at $\mathbf{k}=(\pi,\pi)$ [solid lines in Fig.~\ref{FigFullSpectra}(b)] gains spectral weight and remains visible as a well-defined peak beyond $T \gtrsim 1.4 J$, before thermal broadening dominates above $T > 2 J$. On the other hand, additional weight appears below the magnetic polaron branch, down to frequencies well below the lowest spectral feature at $T=0$~\cite{Bohrdt2020} located at $\mathbf{k}=(\pi/2,\pi/2)$. This indicates that the underlying microscopic processes extract energy from thermal excitations on top of the undoped initial state. At $T \gg J$ the lowest-energy spectral feature we find across the entire BZ is located at $\mathbf{k}=(\pi,\pi)$, which motivates us to study this momentum point more closely.

The additional spectral weight appearing around $\mathbf{k}=(\pi,\pi)$ as temperature is increased has a notable sub-structure. First, at $T=0.71 J$ in Fig.~\ref{FigFullSpectra}(b), we observe one additional low-energy branch emerging $\mathcal{O}(J)$ below the magnetic polaron (following the dashed line). Later, beyond $T \geq 0.99 J$ in Fig.~\ref{FigFullSpectra}(b), a further pronounced maximum emerges between the two previous lines around $\mathbf{k}=(\pi,\pi)$, which remains visible up to high temperatures $T > 4 J$. The same phenomenology, with two additional branches emerging at two different temperatures, is found in our simulations at $t/J=5$, see supplements. In particular, a comparison of the additional branches at $t/J=3$ and $5$ shows that their energy scales as $\propto J$, almost completely independent of $t/J$ in the regime where $t > J$, see supplements for details.

We expect that the existence of such well-defined peaks in the spectrum, appearing at elevated temperatures, is likely related to the small circumference, $L_y=4$, of the cylinder to which our simulations are constrained due to technical limitations. This view is supported by our following theoretical interpretation, and suggests that a less structured spectral feature with a width of several $J$ can be expected below a well-defined magnetic polaron branch in an extended two-dimensional system.

\emph{Nearly-deconfined spinons.---}
To explain the low-energy spectral features emerging around $\mathbf{k}=(\pi,\pi)$ at intermediate temperatures, $0.5 J \lesssim T \lesssim 2 J$, we model the undoped parent AFM as a resonating valence-bond (RVB) state perturbed by a staggered magnetic field representing the non-zero N\'eel order parameter of the AFM ground state~\cite{Lee1988,Piazza2015}. To this end, spin operators are represented by fermionic spinons $\hat{f}_{\mathbf{j},\alpha}$ in the following way, $\hat{\mathbf{S}}_{\mathbf{j}}^\mu= \frac{1}{2} \sum_{\alpha,\beta=\uparrow,\downarrow} \hat{f}_{\mathbf{j},\alpha}^\dagger \sigma^\mu_{\alpha,\beta} \hat{f}_{\mathbf{j},\beta}$, where $\sigma^\mu$ denotes Pauli matrices ($\mu=x,y,z$) and subject to the constraint $\sum_\alpha \hat{f}_{\mathbf{j},\alpha}^\dagger \hat{f}_{\mathbf{j},\alpha} = 1$. A mean-field Hamiltonian for spinons yielding accurate ground state energies (below $1\%$ error) after Gutzwiller projection~\cite{Lee1988,Piazza2015,Trivedi1989} includes a staggered magnetic flux $\pm \Phi$ and a staggered magnetic field $B_{\rm st}$ breaking the ${\rm SU}(2)$ symmetry. It leads to a two-band spinon dispersion $\omega^\pm_{\rm s}(\mathbf{k})$ for $\mathbf{k}$ in the magnetic BZ, with bandwidth $\mathcal{O}(J)$, of which the lower band $\omega^-_{\rm s}(\mathbf{k})<0$ is filled and the upper band $\omega^+_{\rm s}(\mathbf{k})=-\omega^-_{\rm s}(\mathbf{k})$ is empty. 

\begin{figure}[t]
\centering
\includegraphics[width = 0.49\textwidth]{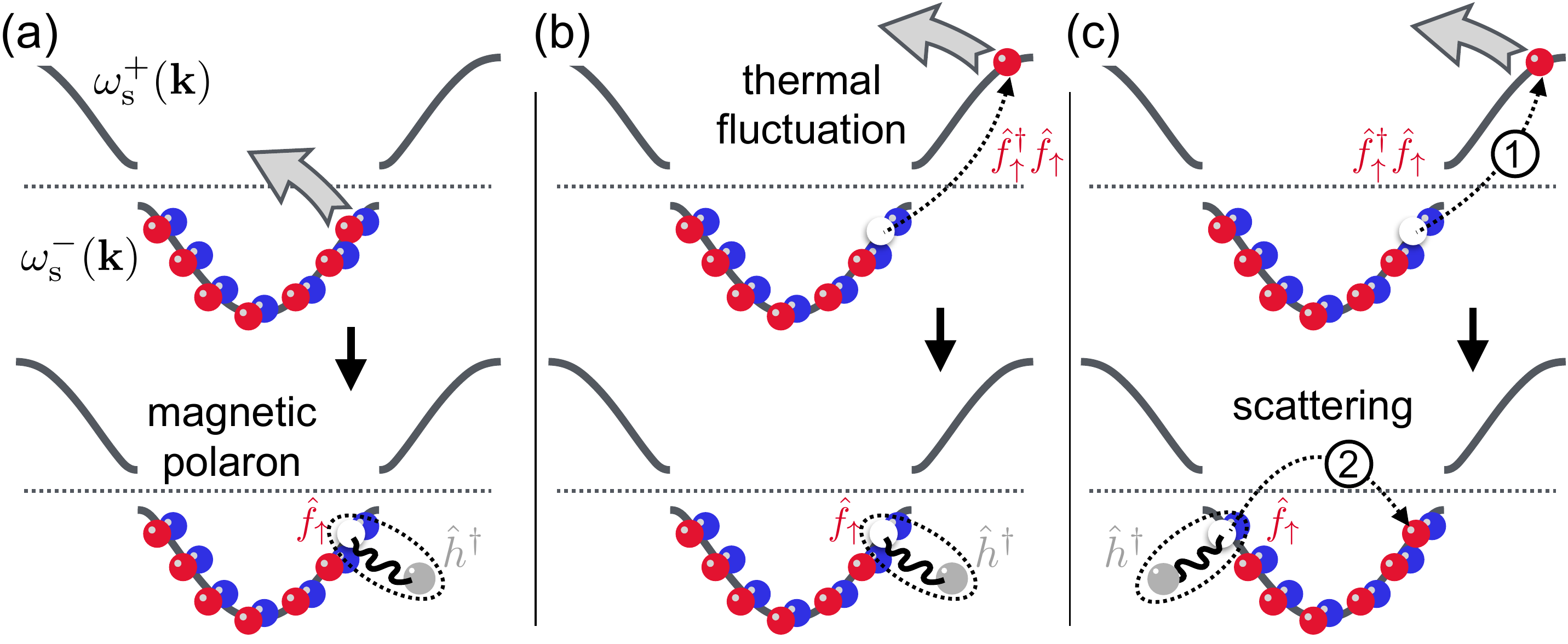}
\caption{Illustration of spinon states contributing to the single-hole ARPES spectrum. Initial states (top row) connected to final states (bottowm row) by $\hat{c}_\sigma = \hat{f}_\sigma \hat{h}^\dagger$ are shown. (a) At $T=0$ a spinon from the lower band is removed and a spinon-chargon ($\hat{f}\hat{h}^\dagger$) bound state constituting the magnetic polaron is formed. (b) At $T>0$, thermal fluctuations lead to spinons in the upper band which can be de-excited by the ARPES beam. (c) Combining scenario (b) with an additional scattering of the spinon-hole $\hat{f}$ off the chargon $\hat{h}^\dagger$ leads to additional spectral features.}
\label{FigSpiononDisp}
\end{figure}

While spinons are confined in the parent AFM, appearing as spin-one pairs (magnons) in the dynamical structure factor, their nearly-deconfined nature has been argued to manifest in high-energy spin-excitations indirectly revealing the structure of weakly interacting  spinons~\cite{Piazza2015}. In addition, it has been shown that individual doped holes $\hat{c}_\sigma \sim \hat{f}_\sigma \hat{h}^\dagger$ can be described as bound states of confined chargons $\hat{h}$ and spinons~\cite{Grusdt2019PRB}. Since chargons are light (mass $\propto 1/t$) and spinons heavy (mass $\propto 1/J$), the dispersion relation of the resulting magnetic polaron is essentially determined by $-\omega_{\rm s}^-(\mathbf{k}) = \omega_{\rm s}^+(\mathbf{k})$ up to an overall normalization~\cite{Beran1996}. Inspection of the spinon dispersion $\omega_{\rm s}^\pm(\mathbf{k})$, see supplements, reveals that this picture correctly reproduces the shape of the magnetic polaron dispersion $\omega_{\rm mp}(\mathbf{k})$~\cite{Grusdt2019PRB}, as confirmed by our fit to the  polaron branch in Fig.~\ref{FigFullSpectra}(b). This picture moreover explains the drop of spectral weight outside the magnetic BZ at low temperatures, see Fig.~\ref{FigSpiononDisp}(a), since the spinon Fermi sea essentially extends to the edge of the magnetic BZ~\cite{Bohrdt2020}.

If a thermally excited spinon $\hat{f}^\dagger_{\mathbf{k}}$ in the upper band, corresponding to $\mathbf{k}$ outside the magnetic BZ~\cite{Bohrdt2020}, exists in the initial state and is de-excited by the ARPES beam, the remaining $\hat{f}_{\mathbf{k}'}$ hole-type spinon excitation in the lower band can form a magnetic polaron bound state at momentum $\mathbf{k}'$ with the chargon. This process is illustrated in Fig.~\ref{FigSpiononDisp}(b) and corresponds to momentum and energy transfers $\mathbf{k}$ and $\omega = - \omega_{\rm s}^+(\mathbf{k}) + \omega_{\rm mp}(\mathbf{k}')$. It readily explains why the lowest-energy spectral weight in Fig.~\ref{FigFullSpectra} appears at $\mathbf{k}=(\pi,\pi)$, where $- \omega_{\rm s}^+(\mathbf{k}) = \omega_{\rm s}^-(\mathbf{k})$ becomes minimal, and for energies $\mathcal{O}(J)$ below $\omega_{\rm mp}^0={\rm min}_{\mathbf{q}} (\omega_{\rm mp}(\mathbf{q}))$. A second type of process allowed for $T>0$ starts from the same thermally excited spinon $\hat{f}^\dagger_{\mathbf{k}}$ in the upper band, which is de-excited by the ARPES beam. In this case it is assumed that the corresponding hole-type spinon excitation $\hat{f}_{\mathbf{k}'}$ in the lower band simultaneously scatters from $\mathbf{k}'$ to a final momentum $\mathbf{k}'+\Delta \mathbf{k}$. This is illustrated in Fig.~\ref{FigSpiononDisp}(c) and implies momentum and energy transfers $\mathbf{k}+\Delta\mathbf{k}$ and $\omega = - \omega_{\rm s}^+(\mathbf{k}) + \omega_{\rm mp}(\mathbf{k}'+\Delta \mathbf{k})$.

The single-hole ARPES spectrum is obtained as a convolution of the initial spinon and the final magnetic polaron states, combining all processes above and implying a broad spectral feature around $\mathbf{k}=(\pi,\pi)$ at finite temperature. However, additional maxima in the spectrum can emerge when taking into account the increased density of states of magnetic polarons at the edge of the magnetic BZ. This effect is particularly pronounced on the four-leg cylinder we study, where low-energy magnetic polarons with a high, quasi-one dimensional density of states along $k_x$ exist only at the inequivalent momenta $(\pi,0)$, $(\pi/2,\pi/2)$, $(0,\pi)$, at similar energy $\omega_{\rm mp}^0$. Allowing only these final magnetic polaron momenta in the processes described in Fig.~\ref{FigSpiononDisp} yields three pronounced peaks observed around $\mathbf{k}=(\pi,\pi)$ in Fig.~\ref{FigFullSpectra}(b): the magnetic polaron at $T=0$ (solid line) and the thermal spinon without (dashed) and with (dotted) scattering. The precise fit parameters we used in Fig.~\ref{FigFullSpectra}(b) are provided in the supplements. The expected features are in good qualitative agreement with our numerically obtained spectra.

\emph{Discussion and Outlook.---}
The $t$-$J$ model studied in this work, and the closely related Hubbard model, have been realized using ultracold atoms in optical lattices~\cite{Esslinger2010,Tarruell2018,Bohrdt2021PWA,bohrdt2024microscopy,Carroll2024} as well as in Rydberg tweezer arrays~\cite{Qiao2025}, allowing for measurements of the single-hole ARPES spectrum~\cite{Brown2019,Bohrdt2018,nielsen2025arXiv}. The robustness of the quasiparticle peak we find in our spectra, up to $T \approx 2 J$, i.e. well above the temperatures reached in state-of-the-art experiments~\cite{Xu2025arXiv}, indicates that spectroscopic studies of magnetic polarons are well within reach. 

A central result of our work is the prediction that low-energy spectral weight -- notably absent at low temperatures -- re-appears outside the magnetic BZ as $T$ increases. We explain this phenomenon by nearly-deconfined spinon excitations in the undoped Heisenberg AFM, which may also underlie Fermi-arc formation. We propose to test this hypothesis further by performing measurements at finite doping, or using pump-probe variants of ARPES~\cite{Schuckert2021a}. By establishing how spectral weight re-appears outside the magnetic BZ we hope to reveal the mechanism leading to its suppression at low $T$ in the first place.

Possible extensions of our work include the study of Hubbard-Mott excitons~\cite{Huang2023,bohrdt2024arXiv}, rotational one-hole excitations~\cite{Grusdt2018Parton,Bohrdt2021PRL} or two-hole ARPES spectra~\cite{Grusdt2023,Bohrdt2023} at finite temperatures. All these approaches will lead to a better microscopic understanding of the emergent charge carriers of doped AFM Mott insulators in the strongly correlated regime, believed to constitute key ingredients of high-$T_c$ superconductivity in cuprate compounds.

\emph{Acknowledgements.---}
The authors thank Sebastian Paeckel and Pit Bermes for helpful discussions. This work was funded in part by the Deutsche Forschungsgemeinschaft under Germany's Excellence Strategy EXC-2111 (Project No.\ 390814868). It is part of the  Munich Quantum Valley, supported by the Bavarian state government with funds from the Hightech Agenda Bayern Plus. F.G. acknowledges funding from the European Research Council (ERC) under the European Union’s Horizon 2020 research and innovation programm (Grant Agreement no 948141) — ERC Starting Grant SimUcQuam.

\section{Supplementary Material}

\begin{figure}[htbp]
    \centering
    \subfloat[]{%
        \includegraphics[width=0.47\textwidth]{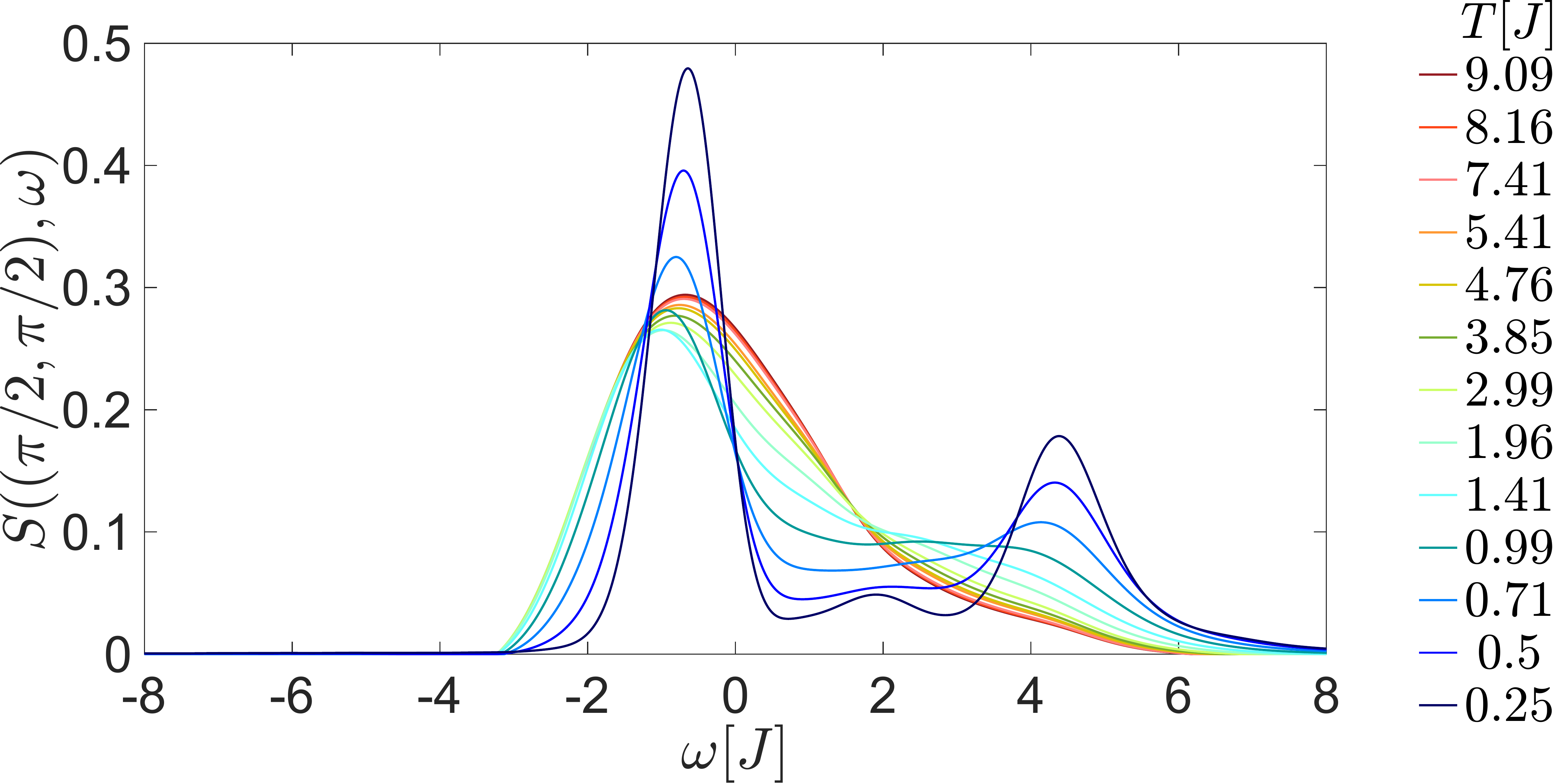}%
        \label{fig:app:spectr_cut_t_1_j_1}%
    }
    \hfill
    \subfloat[]{%
        \includegraphics[width=0.47\textwidth]{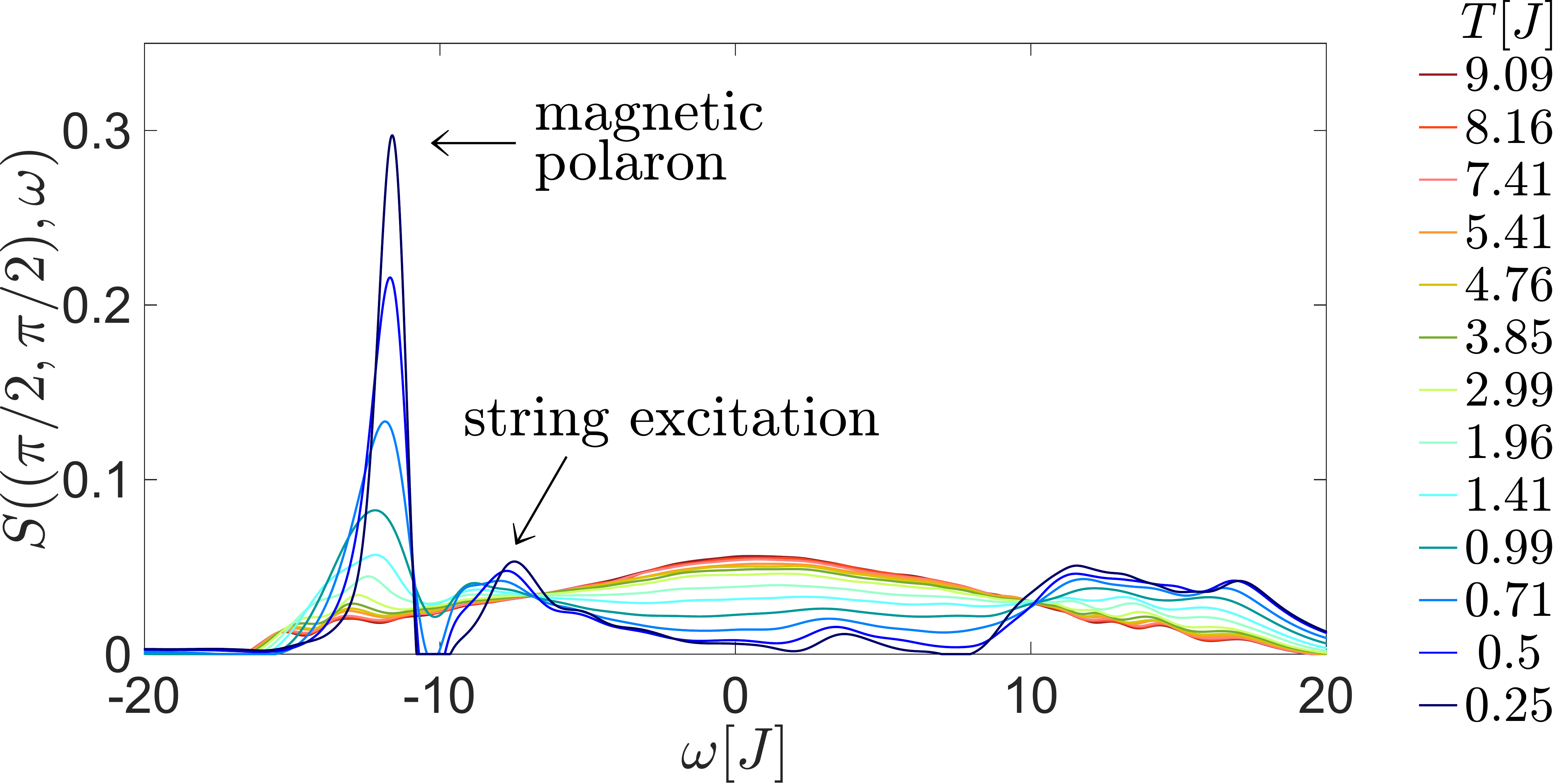}%
        \label{fig:app:spectr_cut_t_5_j_1}%
    }
    \caption{Single-hole ARPES spectrum at the nodal point $\mathbf{k}=(\pi/2,\pi/2)$, computed for various temperatures $T$ indicated on the right. The results are displayed as in \Fig{FigSpectralCut} of the main text, but now for $t/J=1$ in (a) and $t/J=5$ in (b). As in the case $t/J=3$ shown in the main text a well defined magnetic polaron and string excitation peak can be observed in (b).}
    \label{figapp_spectral_cut}
\end{figure}

\begin{figure*}[t]
\centering
\includegraphics[width = \textwidth]{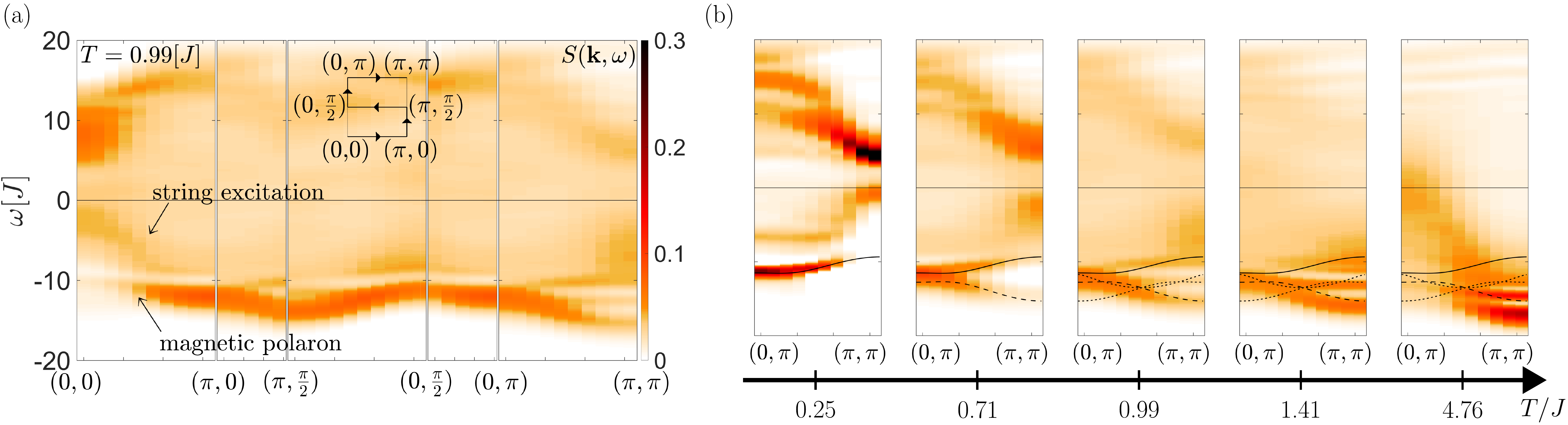}
\caption{Finite temperature ARPES spectra calculated in  the same manner as Fig.~\textcolor{blue}{2} in the main text, but now for $t/J=5$. The fit parameters used here are: $\omega_{\mathrm{MP}}^0=-11.5J$, $\lambda=0.85$ and $J_{\mathrm{eff}}=1J$.}
\label{Figapp_temp_dep_t_5_j_1}
\end{figure*}

\begin{figure*}[t]
\centering
\includegraphics[width = \textwidth]{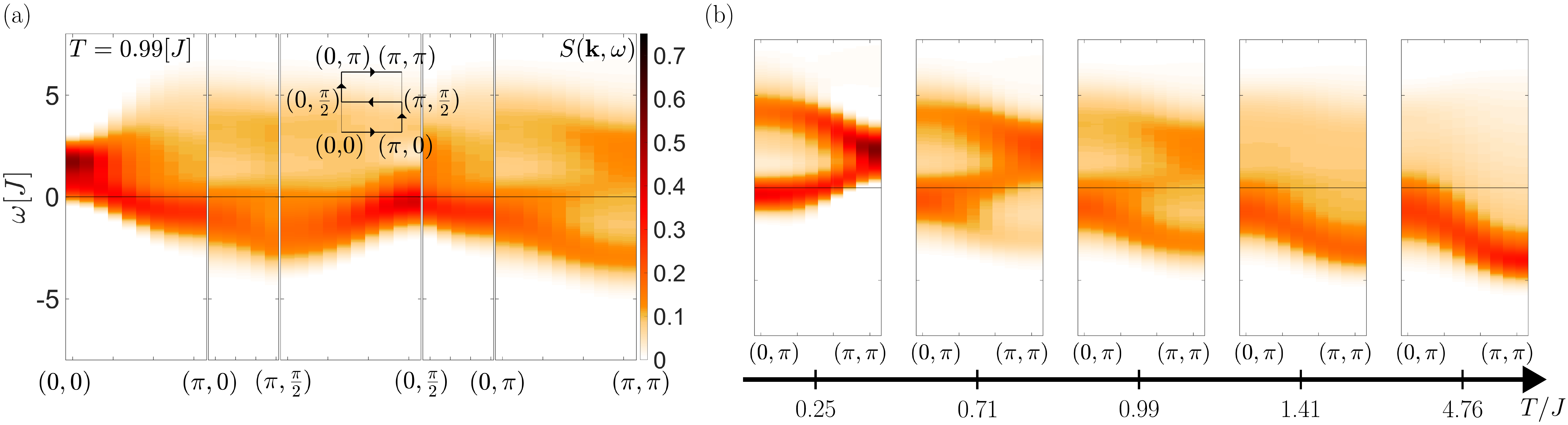}
\caption{Finite temperature ARPES spectra calculated in  the same manner as Fig.~\textcolor{blue}{2} in the main text, but now for $t/J=1$.}
\label{Figapp_temp_dep_t_1_j_1}
\end{figure*}

\emph{Numerical results for other values of $t/J$.---}
In the main text, we focused on the experimentally most relevant case when $t/J=3$. Here we repeat our simulations for ratios $t/J=1$ and $t/J=5$ which provides valuable insights into the dependence of various features in the one-hole spectra on $t$ and $J$, respectively. In \Fig{figapp_spectral_cut} we display the single-hole ARPES spectrum at the nodal point $\mathbf{k}=(\pi/2,\pi/2)$ as in Fig.~\textcolor{blue}{1} of the main text, but now for $t/J=1$ and $5$. In \Fig{Figapp_temp_dep_t_5_j_1} and \Fig{Figapp_temp_dep_t_1_j_1} we present the finite temperature ARPES spectra calculated in  the same manner as Fig.~\textcolor{blue}{2} in the main text, but now for $t/J=1$ and $5$. Especially in the case when $t/J=5$, well defined magnetic polaron and string excitation peaks can be observed, similar to the case $t/J=3$ shown in the main text.

\emph{Theoretical model of nearly-deconfined spinons.---}
In the main text we described the mean-field RVB description of the two-dimensional Heisenberg AFM in terms of spinons. Their dispersion relation can be well approximated as
\begin{equation}
    \omega^\pm_{\rm s}(\mathbf{k})= \pm \sqrt{4 J_{\rm eff}^2 \left| \cos(k_x) e^{-i\frac{\Phi}{4}} + \cos(k_y) e^{i\frac{\Phi}{4}} \right|^2 + \frac{B_{\rm st}^2}{4}},
    \label{eqOmegaSdef}
\end{equation}
where $J_{\rm eff} = \mathcal{O}(J)$ is a fit parameter and the optimal variational values are $\Phi=0.4 \pi$ and $B_{\rm st}=0.44 J_{\rm eff}$~\cite{Piazza2015}.

To understand how the single-hole ARPES spectrum is related to spinons, we use the Lehmann representation
\begin{equation}
    A(\mathbf{k},\omega) = \sum_{m,n} \frac{e^{-\beta E_n}}{Z_0} \delta(\omega - E_m^{\rm 1h} + E_n) \mathcal{M}_{m,n}(\mathbf{k}),
\end{equation}
where $\mathcal{M}_{m,n}(\mathbf{k})=\sum_\sigma |\bra{\psi^{\rm 1h}_m} \hat{c}_{\mathbf{k},\sigma} \ket{\psi_n}|^2$, indices $m$ and $n$ label one and zero hole eigenstates and energies respectively and $\beta=1/(k_{\rm B} T)$; $Z_0$ provides normalization. Hence energy and momentum are conserved and $\omega$ and $\mathbf{k}$ correspond to the difference in energy and momentum of the final and initial states connected by $\hat{c}_{\mathbf{k},\sigma}$. We proceed by decomposing $\hat{c}_{\mathbf{k},\sigma} \propto \sum_{\mathbf{q}} \hat{f}_{\mathbf{k}+\mathbf{q},\sigma} \hat{h}^\dagger_{\mathbf{q}}$, such that $\hat{c}_{\mathbf{k},\sigma}$ creates a spinon-chargon pair. Hence the simplest final state $\ket{\psi^{\rm 1h}_m}$ corresponds to an $\hat{f}_{\mathbf{k}}$-hole excitation with energy $\propto -\omega_{\rm s}^-(\mathbf{k})$ in the band insulator formed by spinons, carrying momentum $\mathbf{k}$ within the magnetic BZ, bound to the light chargon. As discussed in the main text, this state constitutes the magnetic polaron, and similarly the other processes illustrated in Fig.~\textcolor{blue}{3} of the main text can be formally understood.

As explained in the main text, we performed a fit to the one-hole spectra along the cut $(0,\pi)$ to $(\pi,\pi)$, shown by solid, dashed and dotted lines in Fig.~\textcolor{blue}{2}(b). Since the dispersion relation of the magnetic polaron (solid line) is essentially determined by $-\omega_{\rm s}^-(\mathbf{k}) = \omega_{\rm s}^+(\mathbf{k})$, we use the fit 
\begin{equation}
\omega_{\mathrm{MP}}(k_x)=\omega_{\mathrm{MP}}^0 + \lambda [\omega_\mathrm{s}(k_x,k_y=\pi)-\omega_\mathrm{s}(0,\pi)],
\end{equation}
where $\omega_{\mathrm{MP}}^0$ and $\lambda$ are fit parameters.

The process associated with de-excitation of the thermal spinon without scattering (dashed line) can be fitted with 
\begin{equation}
    \omega(k_x)=\omega_{\mathrm{MP}}^0 -  \omega_\mathrm{s}(k_x,k_y=\pi).
\end{equation}
Considering that the resulting magnetic polaron can only exist at the inequivalent momenta $(\pi,0)$, $(\pi/2,\pi/2)$, $(0,\pi)$ de-excitation of the thermal spinon with scattering (dotted line) can be described as
\begin{equation}
\omega(k_x) = 
    \begin{cases}
    \omega_{\mathrm{MP}}^0 -  \omega_\mathrm{s}(k_x-\frac{\pi}{2},\frac{\pi}{2}) \\
    \omega_{\mathrm{MP}}^0 -  \omega_\mathrm{s}(k_x-\pi,\pi).
    \end{cases}
\end{equation}
For the results presented in the main text, we obtained the following fit parameters: $\omega_{\mathrm{MP}}^0=-5.8J$, $\lambda=0.85$ and $J_{\mathrm{eff}}=1J$.

\begin{figure}[t]
\centering
\includegraphics[width = 0.38\textwidth]{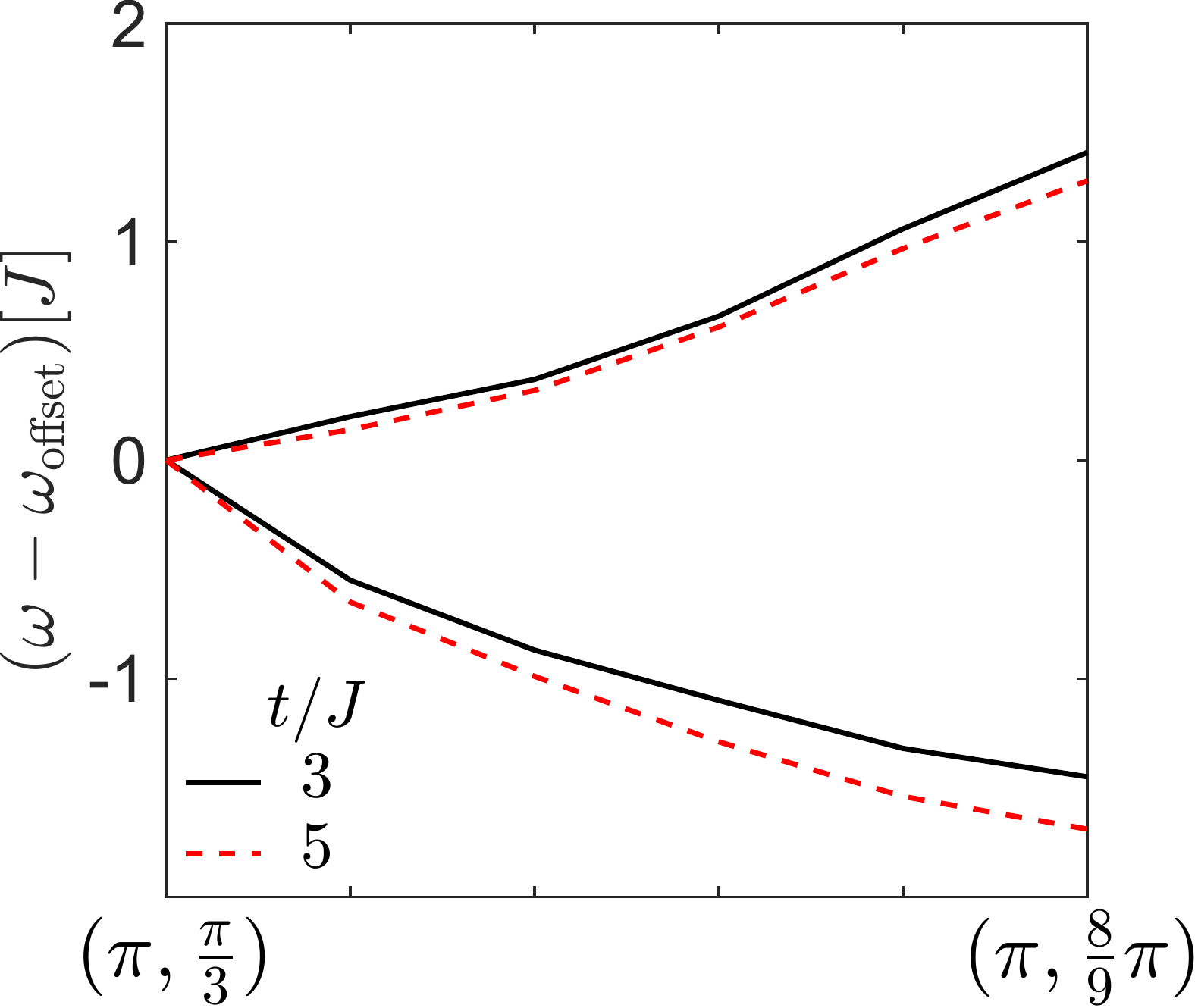}
\caption{Peak locations of the two additional branches in the spectral function extracted at $T=0.71J$ for $t/J = 3$ and $5$. The peak locations have been extracted for the vertical cut $(\pi,0)\xrightarrow{} (\pi,\pi)$. To facilitate comparison between the curves for different $t/J$, the peak locations have been shifted by an offset term $\omega_{\mathrm{offset}}[J]$.}
\label{Figapppeak_locations}
\end{figure}

\emph{Comparison of the additional branches at $t/J=3$ and $5$.---} In \Fig{Figapppeak_locations} we compare the peak locations of the two additional branches in the spectrum at $T=0.71J$ for $t/J = 3$ and $5$. We find that the branches with different $t/J$ are almost positioned on top of each other. As $\omega$ is plotted as a function of $J$, this indicates that their energy scales as $\propto J$ and is nearly independent of $t/J$.

\begin{figure}[t]
\centering
\includegraphics[width = 0.47\textwidth]{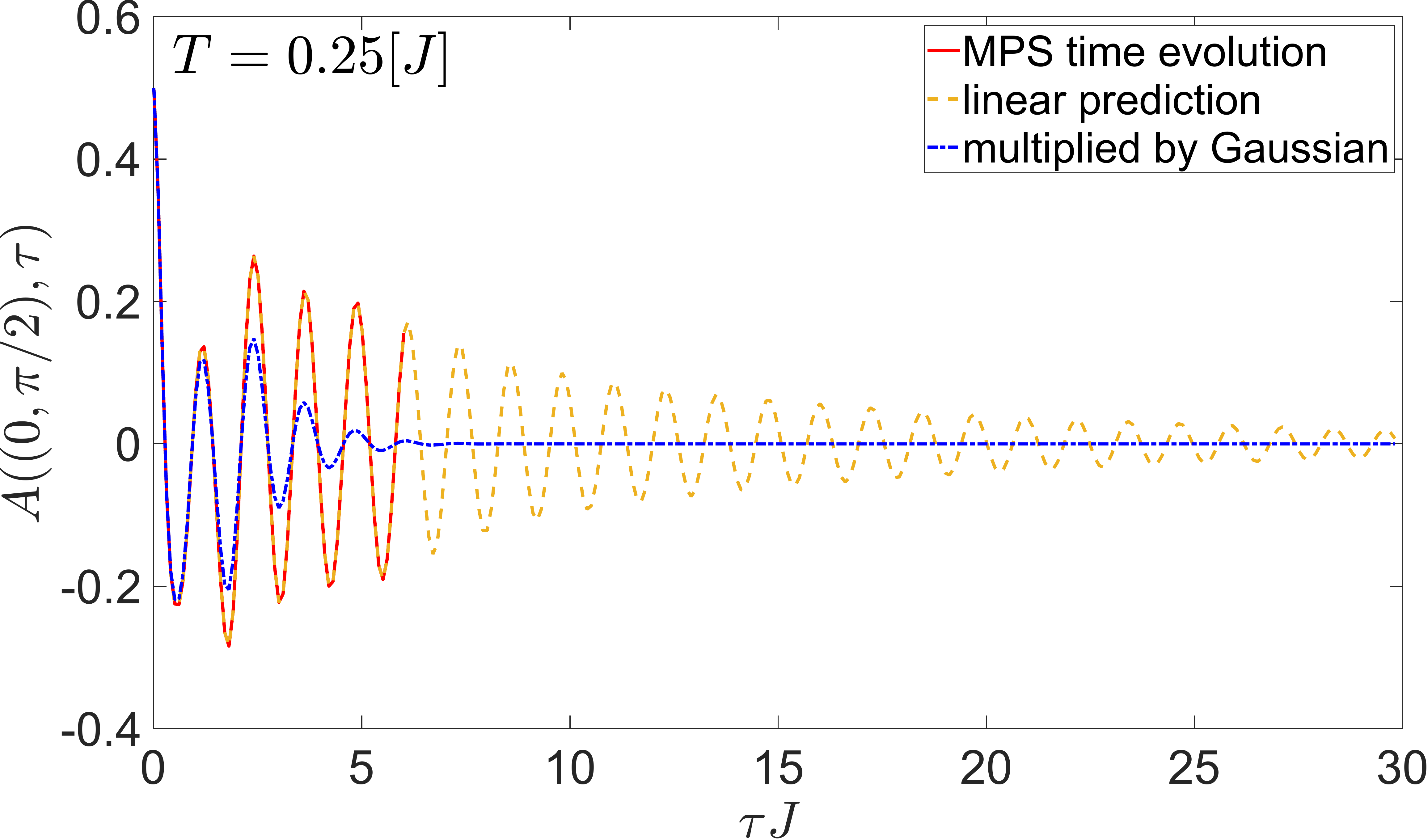}
\caption{Linear prediction and multiplication with Gaussian envelope. After the spatial Fourier transform we use linear prediction to increase the time window (yellow dashed). Prior to the Fourier transform in time $\tau$ we multiply the data with a Gaussian envelope $\propto \me^{-\eta \tau^2}$ of width $\eta=0.1J^2$. The procedure is displayed for $\mathbf{k}=(0,\pi/2)$ and $t/J=3$.}
\label{Figappprediction_damping}
\end{figure}

\emph{Linear prediction and Gaussian envelope.---} In \Fig{Figappprediction_damping} we illustrate linear prediction \cite{leland1989digital} and multiply the time dependent correlation function $A(\mathbf{k},\tau)$ with a Gaussian envelope $\propto \me^{-\eta \tau^2}$  to extend the time window that we can use for the Fourier transform in Eq.~(\textcolor{blue}{2}).

\begin{figure}[t]
\centering
\includegraphics[width = 0.47\textwidth]{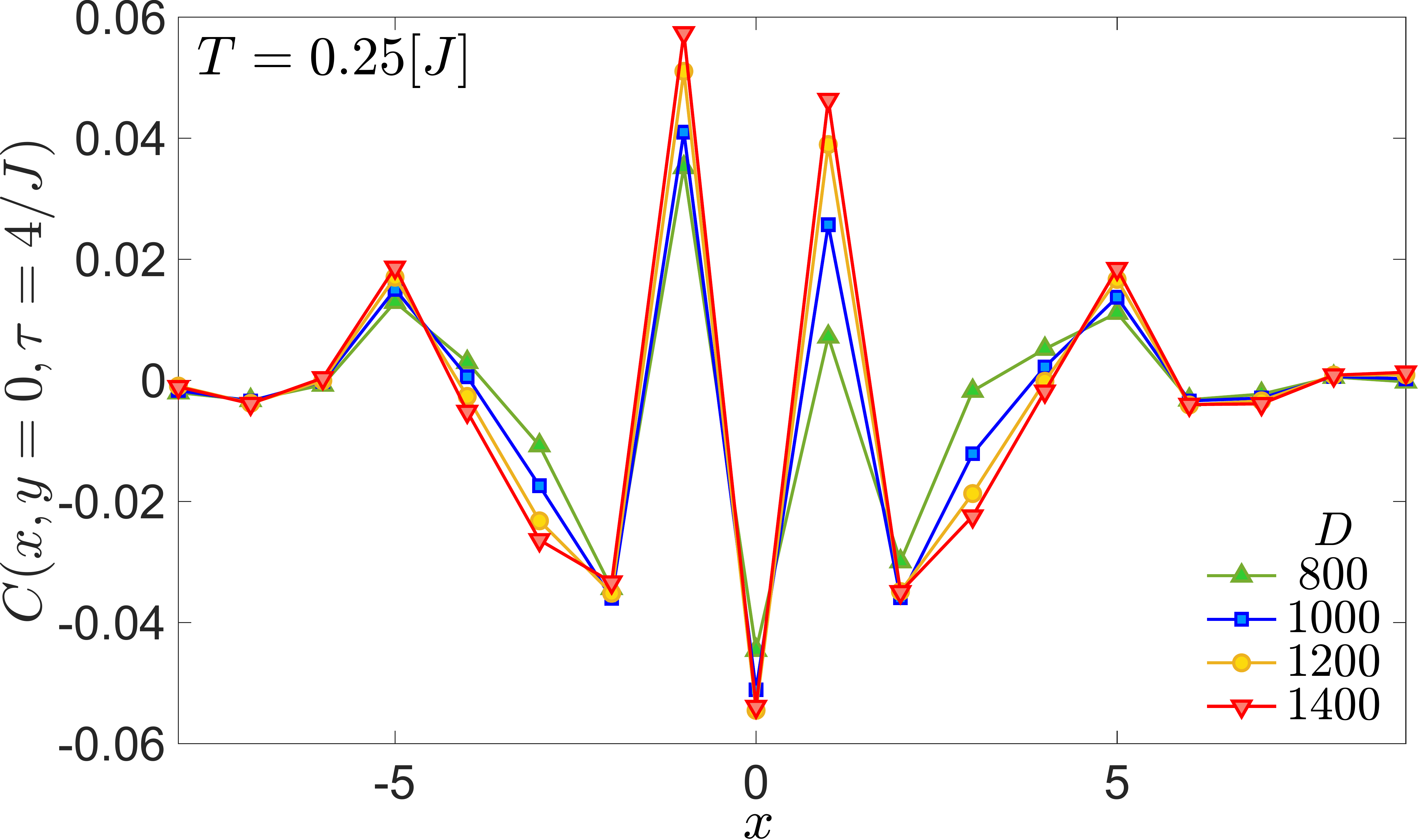}
\caption{Convergence with bond dimension $D$ in the time dependent correlation function $C(x,y=0,\tau)=C_{(x,y),\mathbf{0}}(\tau)$ relative to the origin. The data is displayed for $t/J=3$ at $T=0.25J$. This temperature is the most challenging temperature in our simulation, and the data is displayed at time $\tau = 4/J$. This time represents the point up to which $A(\mathbf{k},\tau)$ multiplied by a Gaussian does still contribute non negligible weight to the Fourier transform in time, see \Fig{Figappprediction_damping}.}
\label{Figappconvergence}
\end{figure}

\emph{Convergence.---} The results presented in the main text have been subjected to rigorous analysis with regard to convergence in several parameters, including the bond dimension $D$, see \Fig{Figappconvergence}.

\emph{Symmetrization.---} The vertical cuts in Fig.~\textcolor{blue}{2} were obtained via the use of symmetrization \cite{Bohrdt2020}. Prior to Fourier transformation, the array of the time-dependent correlation data, which is of size $L_x \times L_y$, is reshaped into an array of size $L_y \times L_x$. As a consequence of the reshaping of the array, the momenta transform as  
\begin{align}
    &k_x \xrightarrow{} k_y \\
    &k_y \xrightarrow{} k_x,
\end{align}  
effectively resulting in the Fourier transformation being unaltered by the reshaping.
It is important to note that this procedure yields an enhanced resolution in the $k_y$ direction.

\FloatBarrier

\nocite{*}
\bibliography{main}

\end{document}